\documentclass[pra,twocolumn,showpacs,superscriptaddress,amsfonts,amsmath,floatfix]{revtex4}

\usepackage{graphicx}

\usepackage[usenames]{color}

\begin{document}

\title{Anderson localization of a Tonks-Girardeau gas in 
potentials with controlled disorder}

\author{J. Radi\'{c}}
\affiliation{Department of Physics, University of Zagreb, PP 332, 10000 Zagreb, Croatia}
\author{V. Ba\v{c}i\'{c}}
\affiliation{Department of Physics, University of Zagreb, PP 332, 10000 Zagreb, Croatia}
\author{D.~Juki\'{c}}
\affiliation{Department of Physics, University of Zagreb, PP 332, 10000 Zagreb, Croatia}
\author{M.~Segev}
\affiliation{Technion, Israel Institute of Technology, Haifa, Israel}
\author{H. Buljan}
\affiliation{Department of Physics, University of Zagreb, PP 332, 10000 Zagreb, Croatia}


\date{\today}

\begin{abstract}
We theoretically demonstrate features of Anderson localization in the Tonks-Girardeau 
gas confined in one-dimensional (1D) potentials with controlled disorder. 
That is, we investigate the evolution of the single particle density and 
correlations of a Tonks-Girardeau wave packet in such disordered potentials. The wave packet 
is initially trapped, the trap is suddenly turned off, and after some time the 
system evolves into a localized steady state due to Anderson localization. 
The density tails of the steady state decay exponentially, while the coherence in 
these tails increases. The latter phenomenon corresponds to the same effect found 
in incoherent optical solitons. 
\end{abstract}

\pacs{05.30.-d, 03.75.Kk, 67.85.De}
\maketitle


\section{Introduction}

The phenomenon of Anderson localization \cite{Anderson1958}, which was originally 
theoretically predicted in the context of condensed matter physics, has been 
experimentally demonstrated in other wave systems including 
optical waves \cite{Wiersma1997,Chabanov2000,Storzer2006,Schwartz2007,Lahini2008} and 
ultracold atomic gases (matter waves) \cite{Billy2008,Roati2008}. 
In the context of Bose-Einstein condensates (BECs), Anderson localization was 
obtained by placing ultracold atomic BECs in elongated, essentially one-dimensional 
disordered \cite{Billy2008} and quasiperiodic incommensurate potentials \cite{Roati2008}, 
which were created optically (see Ref. \cite{Sanchez-Palencia2010} for a recent 
review of the topic). 
The matter waves utilized in those experiments were condensates, i.e., they were 
spatially coherent in the sense that their one-body density matrix factorizes 
$\rho(x_1,x_2)\approx \Phi^*(x_1)\Phi(x_2)$, where $\Phi(x)$ is the condensate wave function. 
However, in reality interactions and/or the presence of the thermal cloud 
affects the spatial coherence in the system. Naturally, the spatial coherence in the 
system is expected to have important implications on localization phenomena, 
since the phenomenon of Anderson localization is deeply connected to interference 
of multiple reflected waves. 
This motivates us to study Anderson localization in a Tonks-Girardeau gas, which 
is a relatively simple example of partially-spatially-coherent Bose gas (i.e., it is not 
condensed).

The Tonks-Girardeau model describes a system of strongly repulsive ("impenetrable") 
bosons, confined in one-dimensional (1D) geometry \cite{Girardeau1960}. Exact solutions 
of the model are found by employing the Fermi-Bose mapping \cite{Girardeau1960,
Girardeau2000}, wherein the Tonks-Girardeau wave function (for both the stationary and 
the time-dependent problems) is constructed from a wave function describing noninteracting 
spinless fermions. In Ref. \cite{Olshanii98} it was suggested that the Tonks-Girardeau 
model can be experimentally realized with ultracold atoms in effectively 1D atomic 
waveguides. This regime is reached at low temperatures, for sufficiently tight 
transverse confinement, and with strong effective interactions
\cite{Olshanii98,Petrov2000,Dunjko2001}. Indeed, in 2004 two groups have 
experimentally realized the Tonks-Girardeau gas \cite{Kinoshita2004,Paredes2004}. 
Furthermore, nonequilibrium dynamics of a 1D Bose gas (including the 
Tonks-Girardeau regime) has been experimentally addressed in the context of 
relaxation to equilibrium \cite{Kinoshita2006}. 
It is known that ground states of the Tonks-Girardeau gas on the ring \cite{Lenard1964}, 
or in a harmonic potential \cite{Forrester2003} are not condensates, because the 
population of the leading natural orbital scales as $\sqrt{N}$, where $N$ is 
the number of particles. Thus, the Tonks-Girardeau gas is only partially spatially 
coherent. The free expansion of the Tonks-Girardeau gas from some initial state 
has been of great interest over the past few years \cite{Rigol2005exp,Minguzzi2005,
DelCampo2006,Gangardt2007}; this type of scenario, i.e., expansion from an initial 
state which is localized (say by a trapping potential) can be used to address 
Anderson localization \cite{Billy2008}.

The experimental demonstrations of Anderson localization in ultracold atomic 
gases were preceded by theoretical investigations of this topic 
(e.g., see Refs. \cite{Damski2003,Roth2003,Sanchez-Palencia2007}, see also 
Ref. \cite{Sanchez-Palencia2010} and references therein). 
The interplay of disorder (or quasiperiodicity) and interactions in a Bose gas 
(from weakly up to strongly correlated regimes), has been often studied in the 
context of the Bose-Hubbard model \cite{Damski2003,Roth2003,Giamarchi1988,
Fisher1989,Gimperlein2005,deMartino2005,Scarola2006,Rey2006,Horstmann2007,
Roux2008,Deng2008,Roscilde2008,Orso2009}. Within the model, 
a transition from a superfluid to a Bose glass phase has been predicted 
to occur \cite{Giamarchi1988,Fisher1989}. The aforementioned interplay 
has been studied by using versatile methods including 
calculating the energy absorption rate \cite{Orso2009}, momentum distribution 
and correlations \cite{deMartino2005,Deng2008}, and expansion dynamics 
\cite{Horstmann2007,Roux2008}. In the limit of strong repulsion, 
the system can be described by using hard-core bosons on the lattice 
\cite{deMartino2005,Horstmann2007,Orso2009}. For these systems, 
by employing the Jordan-Wigner transformation the bosonic system is 
mapped to that of noninteracting spinless fermions, and 
all one-body observables can be furnished from the one body density matrix 
both in the stationary (e.g., see \cite{Rigol2005}) and out-of-equilibrium 
systems \cite{Rigol2005exp}. 
The ground state properties of the hard-core Bose gas in a random lattice 
have been studied in \cite{deMartino2005}, whereas expansion dynamics 
was considered in \cite{Horstmann2007}; both approaches predict the 
loss of quasi long-range order.

Here we study Anderson localization within the framework of the 
Tonks-Girardeau model \cite{Girardeau1960} in one-dimensional disordered potentials. 
We study the expansion of a Tonks-Girardeau wave packet in a potential with 
controlled disorder. The potential is characterized by its correlation distance 
parameter $\sigma$. At $t=0$, the initial wave packet is
in the ground state of a harmonic trap with frequency $\omega$ (with small disorder 
superimposed upon it), and then the trap is suddenly turned off. After some time, 
we find that the system reaches a steady state characterized by exponentially decaying 
tails of the density. We show that the exponents decrease with the increase 
of $\omega$ and the decrease of $\sigma$ in the investigated parameter span
($\sigma=0.13-0.40$~$\mu$m and $\omega= 5-10$~Hz). 
The one-body density matrix $\rho_B(x,y,t)$ of the steady state, that is its 
amplitude $|\rho_B(0,x,t)|$, decays exponentially on the tails of the localized 
wave packet. However, in the region of these tails the degree of first 
order coherence $|\mu_B(0,x,t)|=|\rho_B(0,x,t)|/\sqrt{\rho_B(0,0,t)\rho_B(x,x,t)}$ 
reaches a plateau. These plateaus are connected to the behavior of the 
single-particle states used to construct the Tonks-Girardeau wave function, 
from which we find that the spatial coherence increases in the tails. 
This increase of coherence in the tails has its counterpart 
in incoherent optical solitons \cite{IncOpt}, a phenomenon well understood 
in terms of the modal theory for incoherent light \cite{IncOpt}. 

The paper is organized as follows: In Section \ref{Sec:TGmodel} we 
describe the Tonks-Girardeau model. In Section \ref{Sec:AL} we present our 
numerical results on Anderson localization in this system. 
Finally, in Section \ref{Sec:Conc} we outline our conclusions.

\section{Tonks-Girardeau model}
\label{Sec:TGmodel}

In this section we present the Tonks-Girardeau model which describes 
"impenetrable-core" 1D Bose gas \cite{Girardeau1960,Girardeau2000}. 
We study a system of $N$ identical Bose particles in 1D geometry, 
which experience an external potential $V(x)$. 
The bosons interact with impenetrable pointlike interactions 
\cite{Girardeau1960}, which means that the wave function describing 
the bosons vanishes whenever two particles are in contact, that 
is, $\psi_B(x_1,x_2,\ldots,x_N,t)=0$ if $x_i=x_j$ for any $i\neq j$. 
The wave function $\psi_B$ must also obey the Schr\" odinger equation 
\begin{equation}
i \frac{\partial \psi_B}{\partial t}=
\sum_{j=1}^{N} \left[ -\frac{\partial^2 }{\partial x_j^2}
+ V(x_j) \right] \psi_B;
\label{uncoupled}
\end{equation}
here we use dimensionless units as in Ref. \cite{Buljan2006}, i.e., $x=X/X_0$, 
$t=T/T_0$, and $V(x)=U(X)/E_0$, where $X$ and $T$ are space and time variables 
in physical units, and $U(X)$ is the potential in physical units. 
Given the particle mass $m$, the time scale $T_0=2mX_0^2/\hbar$ 
and energy-scale $E_0=\hbar^2/(2mX_0^2)$ are set by choosing an arbitrary spatial 
length-scale $X_0$. For example, in our calculations below $X_0=1\ \mu$m, while 
the mass corresponds to $^{87}$Rb, which yields the temporal scale
$T_0=2.76\times10^{-3}\ s$, 
and the energy scale $E_0=3.82\times10^{-32}\ J$. The wave functions are normalized as 
$\int dx_1\ldots dx_N |\psi_B(x_1,x_2,\ldots,x_N,t)|^2=1$.

The solution of this system may be written in compact form via the 
famous Fermi-Bose mapping, which relates the Tonks-Girardeau bosonic wave function 
$\psi_B$ to an antisymmetric many-body wave function $\psi_F$ 
describing a system of noninteracting spinless fermions 
in 1D \cite{Girardeau1960}:

\begin{equation}
\psi_B = \Pi_{1\leq i < j\leq N} \mbox{sgn}(x_i-x_j) \psi_F(x_1,x_2,\ldots,x_N,t).
\label{mapFB}
\end{equation}
In many physically relevant situations, the fermionic wave function
$\psi_F$ can be written in a form of the Slater determinant,
\begin{equation}
\psi_F(x_1,\ldots,x_N,t)=
\frac{1}{\sqrt{N!}} \det_{m,j=1}^{N} [\psi_m(x_j,t)], 
\label{psiF}
\end{equation} 
where $\psi_m(x,t)$ denote $N$ orthonormal single-particle 
wave functions obeying a set of uncoupled single-particle 
Schr\" odinger equations 
\begin{equation}
i\frac{\partial \psi_m}{\partial t}=
\left [ - \frac{\partial^2 }{\partial x^2}+
V(x) \right ] \psi_m(x,t), \ m=1,\ldots,N.
\label{master}
\end{equation}
In Eqs. (\ref{mapFB})-(\ref{master}) we have outlined the construction 
of the many-body wave function describing the Tonks-Girardeau gas in an external 
potential $V(x)$, both in the static \cite{Girardeau1960} and 
time-dependent case \cite{Girardeau2000}. 

Given the wave function $\psi_B$, we can straightforwardly calculate 
all one-body observables furnished by the reduced single-particle density 
matrix (RSPDM), 
\begin{eqnarray}
\rho_{B}(x,y,t) & = &  N \int \!\! dx_2\ldots dx_N \, \psi_B(x,x_2,\ldots,x_N,t)^*
\nonumber \\
&& \times \psi_B(y,x_2,\ldots,x_N,t);
\end{eqnarray} 
by employing the formalism presented in Ref. \cite{Pezer2007}. 
If the RSPDM is expressed in terms of the single-particle wave functions $\psi_m$ as 
\begin{equation}
\rho_{B}(x,y,t)=\sum_{i,j=1}^{N}
\psi^{*}_{i}(x,t)A_{ij}(x,y,t)\psi_{j}(y,t),
\label{expansion}
\end{equation}
it can be shown that the $N\times N$ matrix ${\mathbf A}(x,y,t)=\{ A_{ij}(x,y,t) \}$ 
has the form
\begin{equation}
{\mathbf A}(x,y,t)=  ({\mathbf P}^{-1})^{T} \det {\mathbf P},
\label{formulA}
\end{equation} 
where the entries of the matrix ${\mathbf P}$ are 
$P_{ij}(x,y,t)=\delta_{ij}-2\int_{x}^{y}dx' \psi_{i}^{*}(x',t)\psi_{j}(x',t)$
($x<y$ without loss of generality) \cite{Pezer2007}.

\section{Numerical results on Anderson localization in a Tonks-Girardeau gas}
\label{Sec:AL}

In order to investigate Anderson localization of the Tonks-Girardeau gas, 
we perform numerical simulations designed in the fashion of optical \cite{Schwartz2007}
and matter wave \cite{Billy2008} experiments which were conducted recently 
to demonstrate Anderson localization. 
We investigate dynamics of a Tonks-Girardeau wave packet in a disordered potential $V_D$(x), 
where the initial wave packet (at $t=0$) is localized in space by some trapping potential. 
After long time of propagation the wave packet reaches some steady state.
Anderson localization is indicated by the exponential decay of the density of 
the wave packet in this steady state. 

More specifically, we assume that initially, at $t=0$, the gas
is in the ground state of the harmonic oscillator potential, with the small 
controlled disordered potential superimposed upon it, that is,
\begin{equation}
V(x)=V_D(x)+\nu^2 x^2 \mbox{ for $t< 0$.}
\end{equation}
At $t=0$ the trapping potential is suddenly turned off, i.e., 
\begin{equation}
V(x)=V_D(x) \mbox{ for $t>0$,}
\end{equation}
after which the density and correlations of the gas begin to evolve. 
This means that at $t=0$ the wave function $\psi_B$ is given by 
Eqs. (\ref{mapFB}) and (\ref{psiF}) where $\psi_m(x,t=0)$ is the 
$m$th single-particle eigenstate of the potential $V_D(x)+\nu^2 x^2$. 
The subsequent evolution of $\psi_m$ is given by Eq. (\ref{master}) where 
the potential is given solely by the disordered term $V(x)=V_D(x)$. 

The construction of the disordered potential $V_D(x)$ used in our numerical simulations 
is described in the Appendix. The disordered potential can be characterized in terms 
of its correlation functions; the autocorrelation function is defined by 
\begin{equation}
A_{C}(x)=\langle \overline{V}_{D}(x'-x) \overline{V}_{D}(x') \rangle_{x'},
\end{equation}
where $\overline{V}_{D}(x)=V_D(x)-\langle V_D(x') \rangle_{x'}$, and 
$\langle \cdots \rangle_{x'}$ denotes a spatial average over $x'$. For the 
disordered potentials in our simulations we have approximately 
\begin{equation}
A_{C}(x)=V_0^2 \frac{\sin^2(x/\sigma)}{(x/\sigma)^2},
\end{equation}
where $\sigma$ denotes the spatial correlation length of the disordered potential, 
whereas $V_0^2=\langle \overline{V}^2_{D}(x) \rangle_{x}$ denotes its amplitude. 
The spatial power spectrum of the potential has support in the interval 
$[-K_{cut},K_{cut}]$, where the cut-off value is $K_{cut}= 2/\sigma$. 
Thus, the potential $V_D(x)$ has the autocorrelation function identical to that of the 
optical speckle potentials used in the experiments, e.g., see \cite{Billy2008}. 

\begin{figure}[h!]
\centerline{
\mbox{\includegraphics[width=0.35\textwidth]{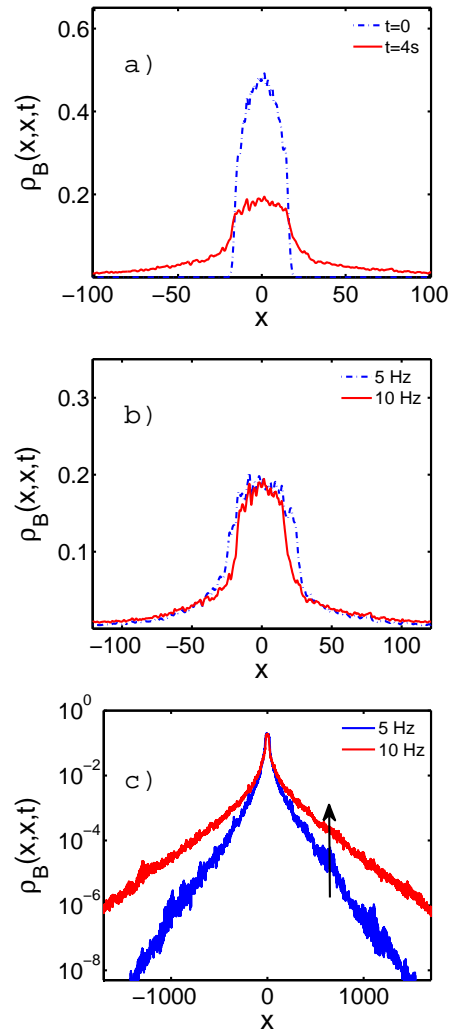}}
}
\caption{
(color online)
Anderson localization in a Tonks-Girardeau gas in dependence 
of the initial trap parameter $\nu$ (i.e., $\omega$). 
The parameters of the disordered potential are ($\sigma=0.13$ and $V_0=0.465$). 
(a) The averaged density of the Tonks-Girardeau wavepacket at $t=0$ and after 
$t=1450$ (=$4$~s) of propagation. The initial state corresponds to 
$\nu=8.67\times10^{-2}$ ($\omega=10$~Hz). 
(b) Shown is the density of a Tonks-Girardeau gas (in the localized steady 
state) after $t=1450$ (=$4$~s) of propagation in a disordered potential.
Blue dot-dashed line corresponds to $\nu=4.34\times10^{-2}$ ($\omega=5$~Hz), whereas
red solid line corresponds to $\nu=8.67\times10^{-2}$ ($\omega=10$~Hz). 
(c) Same as figure (b) on a logarithmic scale. Blue line corresponds to 
($\omega=5$~Hz), and red line corresponds to $\omega=10$~Hz; arrow indicates the 
increase of $\omega$.
For $|x|$ larger than some value (call it $L_{t}$), the density decays 
exponentially, which characterizes Anderson localization.  
The density-tails decay slower for larger initial trap parameter $\omega$ 
(see text for details).
}
\label{gust_nu}
\end{figure}

\begin{figure}
\centerline{
\mbox{\includegraphics[width=0.35\textwidth]{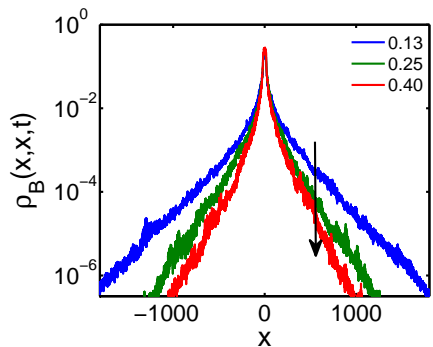}}
}
\caption{
(color online) 
Anderson localization in a Tonks-Girardeau gas in dependence 
of the disorder parameter $\sigma$. 
The averaged density of a Tonks-Girardeau gas after $t=1450$ (=$4$~s) of propagation in 
the disordered potential. 
The plots correspond to $\sigma=0.13$ (blue line), $\sigma=0.25$ (green line), 
and $\sigma=0.40$ (red line); arrow indicates the increase of $\sigma$.
The initial state corresponds to $\omega=10$~Hz, while the amplitude of the 
disordered potential is $V_0\approx 0.47$ (see text for details). 
}
\label{gust_sigma}
\end{figure}

\begin{figure}
\centerline{
\mbox{\includegraphics[width=0.35\textwidth]{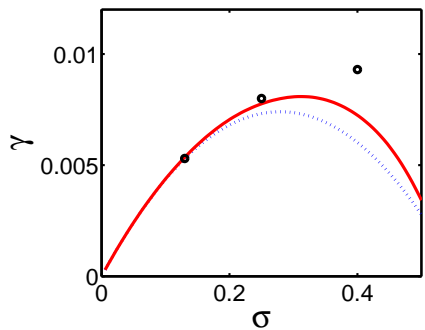}}
}
\caption{
(color online) The exponential decay rates $\gamma^{(2)}$ (dotted blue line) and 
$\gamma^{(2)}+\gamma^{(3)}$ (solid red line), obtained with perturbation theory 
\cite{Lugan2009}, in dependence of the correlation length 
of the potential $\sigma$. Circles represent decay rates obtained numerically.
The initial trap parameter is $\omega=10$~Hz (see text for details).}
\label{gamma_sigma}
\end{figure}

The asymptotic steady state of the system depends on the parameters of the disordered 
potential $\sigma$ and $V_0$, and on the initial state, that is, the harmonic trap 
parameter $\nu$. In fact, since the dynamics of the Tonks-Girardeau gas 
is governed by a set of uncoupled Schr\" odinger equations, it 
follows from the simple scaling of units outlined below Eq. (\ref{uncoupled}), 
that there are in fact only two independent parameters; 
thus we investigate the dynamics in dependence of $\nu$ and 
$\sigma$, and keep $V_0$ at an approximately constant value. 
We have performed our numerical simulations in a region of the 
parameter space which was accessible with our numerical capabilities, 
but which is relevant to experiments \cite{Billy2008,Roati2008}. 
We have varied the correlation length $\sigma$ of the potential 
from $0.13$ up to $0.40$ (corresponding to $0.13\ \mu$m and $0.40\ \mu$m
since the spatial scale is chosen to be $X_0=1\ \mu$m), and the harmonic trap parameters 
in the interval $\nu=4.34-8.67\times10^{-2}$ (corresponding to $\omega=5-10$ Hz). 
The number of particles used in our simulations is relatively small, $N=13$, 
due to the computer limitations, however, despite of this, one can 
use our simulations to infer general conclusions that would be valid 
in an experiment with larger $N$. 
It should be emphasized that all plots of densities and correlations are 
ensemble averages made over 40 realizations of the disordered potentials.

First we investigate the behavior of the single-particle density. 
In Figure \ref{gust_nu} we show $\rho_B(x,x,t=1450)$ versus $x$ for 
$(\sigma,V_0)=(0.13,0.465)$, and two values of $\nu$: $\nu=4.34\times10^{-2}$ 
($\omega=5$~Hz) and $\nu=8.67\times10^{-2}$ ($\omega=10$~Hz). 
In Fig. \ref{gust_nu}(a) we compare the initial density (at $t=0$), 
with the density at time $t=1450$ ($=4$~s), at which the steady state regime 
is already achieved (all graphs below which describe the steady state 
are also calculated at this time). We observe that the steady-state density 
has a broad central part with a fairly flat top, and decaying tails on its sides. 
The central part is composed of many single-particle states $\psi_j$. 
In Fig. \ref{gust_nu}(b) we plot the steady state density for two values 
of $\omega$. For $\omega=5$~Hz, the central part is broader than for $\omega=10$~Hz, 
but the tails are decaying faster with the increase of $|x|$, as shown in 
Fig. \ref{gust_nu}(c), where the densities are plotted in the logarithmic scale. 
We clearly see that for $|x|$ larger than some value (call it $L_{t}$), 
the density decays exponentially, which indicates Anderson localization. 
We have fitted the tails to the exponential curve 
$\rho_B(x,x,t)\propto \exp(-\Lambda |x|)$ and obtained $\Lambda=0.0097$ for 
$\omega=5$ Hz, and $\Lambda=0.0053$ for $\omega=10$ Hz, that is, we find that 
the density-tails decay slower for larger initial trap parameter $\omega$.
For larger values of $\omega$, the trap is tighter and the initial state has 
larger energy and broader momentum distribution, therefore, it is harder to achieve 
localization of the wave packet (e.g., see \cite{Sanchez-Palencia2007,Lugan2009}). 
Another way to interpret these simulations is in terms of the spatial correlation 
distance of the wave packet. An incoherent wave packet can be characterized by using the spatial 
correlation distance, which determines a spatial degree of coherence; 
this quantity is inversely proportional to the width of the spatial power spectrum. 
If the spatial correlation distance decreases, it is harder to achieve localization. 

\begin{figure}
\centerline{
\mbox{\includegraphics[width=0.35\textwidth]{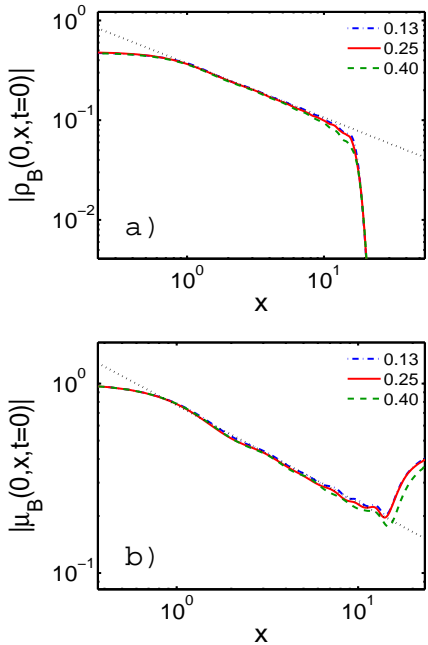}}
}
\caption{
(color online) First-order correlations in the initial state decay algebraically. 
Shown are the single-particle density matrix $|\rho_B(0,x,0)|$ (a), 
and the degree of first-order coherence $|\mu_B(0,x,0)|$ (b), at $t=0$
for the initial state corresponding to $\omega=10$~Hz.
The graphs are plotted for three values of $\sigma$ as indicated in the 
legend. Black dotted lines depict the fitted curves $|\rho_B(0,x,0)|\sim |x|^{-0.54}$ 
and $|\mu_B(0,x,0)|\sim |x|^{-0.51}$.
}
\label{rhot=0}
\end{figure}
\begin{figure}
\centerline{
\mbox{\includegraphics[width=0.35\textwidth]{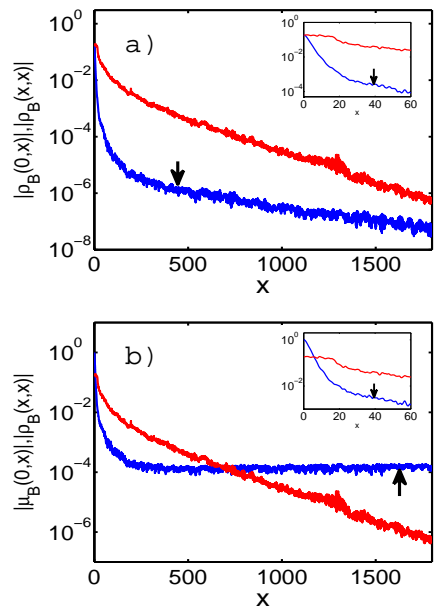}}
}
\caption{
(color online) Correlations in the steady (Anderson localized) state. 
The single-particle density matrix $|\rho_B(0,x,t)|$ [blue line in (a), 
indicated with the arrow], and the degree of first order coherence 
$|\mu_B(0,x,t)|$ [blue line in (b), indicated with the arrow], at the time 
$4$~s. The parameters used are $\sigma=0.13$ and $\omega=10$~Hz. Red lines 
(in both panels) depict the single particle density $\rho_B(x,x,t)$. 
Insets enlarge the region where $|x|$ is small and where correlations 
decay approximately exponentially. For $|x|$ in the region of the density 
tails ($|x|>L_t$), $|\mu_B(0,x,t)|$ reaches a plateau. 
See text for details. 
}
\label{rhot}
\end{figure}
\begin{figure}
\centerline{
\mbox{\includegraphics[width=0.35\textwidth]{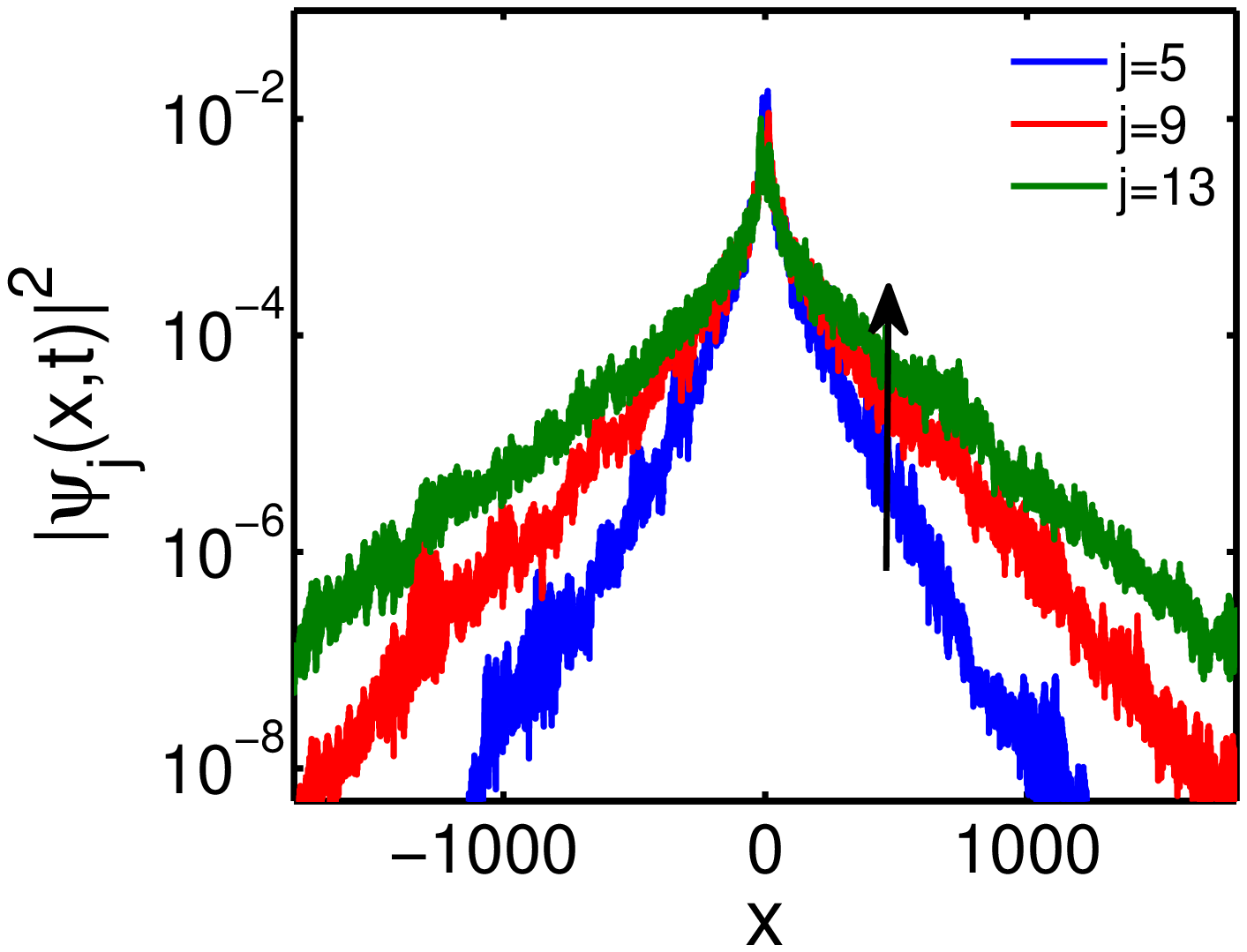}}
}
\caption{
(color online) The single-particle states $|\psi_j(x,t)|^2$ for $j=5,9$, and $13$
in the (Anderson localized) steady state. The arrow indicates increase of $j$. 
The parameters used are $\sigma=0.13$ and $\omega=10$~Hz. 
The single-particle states for larger $j$ (larger in energy) decay slower 
with the increase of $|x|$. See text for details. 
}
\label{figpsij}
\end{figure}

In Fig. \ref{gust_sigma} we display dependence of the density $\rho_B(x,x,t)$ 
versus $x$ for $\nu=8.67\times10^{-2}$ ($\omega=10$~Hz), and three values of $(\sigma,V_0)$: 
$(0.13,0.465)$, $(0.25,0.478)$, and $(0.40,0.485)$. 
Note that $V_0$ can be regarded as a constant close to $0.47$ and we will omit 
to explicitly write the values of $V_0$ besides $\sigma$ in further text; the 
variations of $V_0$ are a consequence of the method utilized to construct the random 
potential. We observe that the exponential tails decay faster for larger values of $\sigma$.

In order to underpin our observations we compare our numerical results to the 
predictions of the formalism presented in Refs. \cite{Sanchez-Palencia2007,Lugan2009}, 
which has been used to study Anderson localization of a Bose-Einstein condensate. 
In the approach of Ref. \cite{Sanchez-Palencia2007}, the wave packet is considered 
to be a superposition of almost independent plane waves ($k$-components); the wave 
packet had a high momentum cut-off at the inverse healing length (of the condensate). 
The assumption that the $k$-components are almost independent means that we should be able to 
employ this formalism here as well, despite of the fact that our wave packet 
is only partially coherent (that is, it is in the Tonks-Girardeau regime, 
rather than being condensed).
However, for our wave packets, the momentum distribution does not have a 
clear cut-off but rather decays smoothly to zero as $k \rightarrow \infty$,
and we must adopt a somewhat different procedure to determine the 
decay rate of the whole wave packet from the decay rate of the $k$-components. 
Within the formalism \cite{Sanchez-Palencia2007,Lugan2009}, the exponential 
decay-rate of every $k$-component is calculated by using perturbation theory, 
and the decay rate $\gamma(k)$ is given as a series $\gamma(k)=\sum_{n\geq 2}\gamma^{(n)}$ 
in increasing orders \cite{Lugan2009}. 
The leading contribution $\gamma^{(2)}$ to the decay rate  $\gamma(k)$ arises from 
the Born approximation, wherein \cite{Sanchez-Palencia2007,Lugan2009,notation} 
\begin{equation}
\gamma^{(2)}(k)=\frac{\pi}{4}  \left( \frac{V_0}{k} \right)^2   \sigma (1-k\, \sigma) 
\end{equation}
for $k<\sigma^{-1}$, and zero otherwise; the next order term is given in 
the Appendix, see also \cite{Lugan2009}. 
In order to evaluate the decay rate of the expanding wave packet from 
$\gamma(k)$ we adopt the following simple procedure. We assume that there 
is an effective high momentum cut-off $k_{hmc}$, which we evaluate as follows:
For the initial wave packets (determined by the trap frequency), we have numerically 
calculated the Lyapunov exponents $\Lambda$ determining localization. 
For the potential parameter $\sigma=0.13$ ($V_0=0.465$), and the initial condition 
corresponding to $\omega=10$~Hz (for these parameters the system is close to the Born 
approximation regime \cite{Sanchez-Palencia2007}), we choose the effective high 
momentum cut off $k_{hmc}$ such that it fits the numerically calculated decay rate, 
that is we extract $k_{hmc}$ from the equation $\Lambda=\gamma(k_{hmc})$;
this yields $k_{hmc}=1.79$. 
Then, by using this value, we calculate the Lyapunov exponents $\gamma(k_{hmc})$ 
in dependence of $\sigma$. In Fig. \ref{gamma_sigma} we illustrate the functional dependence 
$\gamma(k_{hmc})$ vs. $\sigma$; dotted blue line depicts the Born approximation 
$\gamma^{(2)}$, and solid red line depicts $\gamma^{(2)}+\gamma^{(3)}$. We see that 
within the parameter regime studied here the trend is well described with the 
perturbation approach. Quantitative deviations occur because higher order terms 
of the perturbation theory are not negligible and should be taken into account. 

Next we focus on correlations contained within the reduced single-particle 
density matrix $\rho_{B}(x,y,t)$. Suppose that we are interested in the 
phase correlations between the center (at zero) and the rest of the cloud 
(at some $x$-value); the quantity $\rho_{B}(0,x,t)$ will decay to zero with 
the increase of $|x|$ even if the field is perfectly coherent simply because 
the density decays to zero on the tails. 
In order to extract solely correlations from the RSPDM, we observe 
the behavior of the quantity \cite{Naraschewski1999}
\begin{equation}
\mu_B(x,y,t)=\frac{\rho_{B}(x,y,t)}{\sqrt{\rho_{B}(x,x,t)\rho_{B}(y,y,t)}},
\label{ccf}
\end{equation}
which is the degree of first-order coherence \cite{Naraschewski1999}
(in optics it is sometimes referred to as the complex coherence 
factor \cite{MandelWolf}). 
In the context of ultracold gases $\mu_B(x,y,t)$ can be interpreted as follows: 
If two narrow slits were made at points $x$ and $y$ of the 1D 
Tonks-Girardeau gas, and if the gas was allowed to drop from these 
slits, expand and interfere, $\mu_B(x,y,t)$ expresses the modulation depth of the 
interference fringes. 
In this work we investigate correlations between the central point of 
the wave packet and the tails: $\mu_B(0,x,t)$.

In Fig. \ref{rhot=0} we show the averages of the magnitudes of the one-body 
density matrix $|\rho_B(0,x,t)|$, and the degree of first order coherence $|\mu_B(0,x,t)|$, 
at time $t=0$ for the initial state corresponding to $\omega=10$~Hz and
for three values of $\sigma$.
From previous studies of the harmonic potential ground-state (e.g., see Ref. \cite{Forrester2003} 
for the continuous Tonks-Girardeau gas and \cite{Rigol2005} for hard-core bosons 
on the lattice) it follows that in a fairly broad interval of $x$-values, 
both $|\rho_B(0,x,t=0)|$ and $|\mu_B(0,x,t=0)|$ decay approximately as a power law 
$|x|^{-\gamma_0}$ with the exponent $\gamma_0 = 0.5$ \cite{Forrester2003,Rigol2005}, 
despite of the fact that the density is not homogeneous; the density dependent 
factors multiplying the power law are also known \cite{Forrester2003,Rigol2005}. 
We have observed that the initial correlation functions 
are well fitted to the power law: $|\rho_B(0,x,0)|\sim |x|^{-0.54}$ 
and $|\mu_B(0,x,0)|\sim |x|^{-0.51}$ for $\omega=10$~Hz (for $\omega=5$~Hz,
we obtain $|\rho_B(0,x,0)|\sim |x|^{-0.60}$ and $|\mu_B(0,x,0)|\sim |x|^{-0.55}$). 
The power-law decay of correlations indicates presence of 
quasi long-range order. Apparently, the properties of the small random potential 
do not significantly affect the correlations of the initial state for the trap 
strengths $\omega$, and disorder parameters used in our simulations. 
This happens because the initial single particle states are localized by 
the trapping potential, rather than by disorder (their decay is Gaussian). 
The effect of disorder on these states becomes more significant for weaker traps, 
because the disordered potential becomes nonegligible in comparison to the 
harmonic term $\nu^2x^2$ in a broader region of space. 
In fact, we expect that if one keeps the number of particles constant, 
for sufficiently shallow traps, disorder would qualitatively change the behavior 
of the correlations in the initial state, in a similar fashion 
as when the trap is absent. However, probing Anderson localization by using 
transport (i.e., expansion of an initially localized wave packet), is perhaps more 
meaningful for tighter initial traps, where the initial wave packets are 
localized by the trap rather than by disorder.

For very small values of $|x|$, and for very large values (at the very tails 
of the wave packet) there are deviations from the power law behavior 
\cite{Forrester2003,Rigol2005}. 
The behavior of $|\mu_B(0,x,0)|$ at the tails, where $|\mu_B(0,x,0)|$ starts to grow 
up to some constant value is attributed to the fact that higher single-particle 
states $\psi_m(x,0)$ decay at a slower rate with the increase of $|x|$, 
and therefore spatial coherence increases in the tails (see 
also the discussion below). 

\begin{figure}
\centerline{
\mbox{\includegraphics[width=0.35\textwidth]{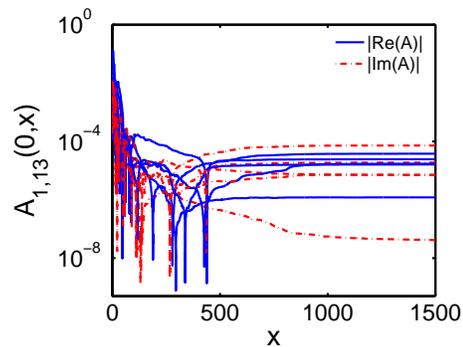}}
}
\caption{
(color online) The absolute value of the real and imaginary part of 
$A_{ij}(0,x,t)$ for $i=1$ and $j=13$, and five different realizations of the 
disordered potential. The parameters used in the simulation 
are $\sigma=0.13$ and $\omega=10$~Hz. For sufficiently large $|x|$, 
$A_{ij}(0,x,t)$ reaches a constant value. See text for details. 
}
\label{Aij}
\end{figure}
\begin{figure}
\centerline{
\mbox{\includegraphics[width=0.35\textwidth]{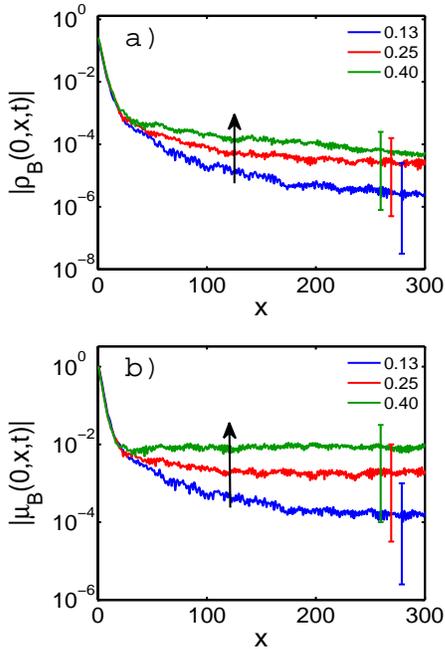}}
}
\caption{
(color online) Correlations in the steady (Anderson localized) state
in dependence of the disorder parameter $\sigma$. Shown are 
the correlations at $t=1450$ ($=4$~s) for $\sigma=0.13$, $0.25,$ and $0.40$. 
The arrows indicate increase of $\sigma$, whereas the vertical bars indicate the 
uncertainty in correlations at the plateau values.
For a given set of parameters $\sigma$ and $\omega$, $90\%$ 
of the simulations (one simulation is made for a single realization 
of $V_D$) fall within the vertical bars. The initial trap 
parameter is $\omega=10$~Hz. See text for details. 
}
\label{corr_sig}
\end{figure}

\begin{figure}
\centerline{
\mbox{\includegraphics[width=0.35\textwidth]{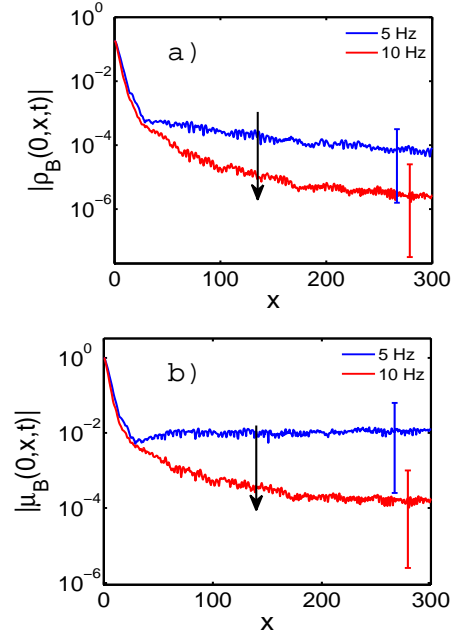}}
}
\caption{
(color online) 
Correlations in the steady state in dependence of the initial trap 
parameter $\omega$. Shown are the correlations at $t=1450$ ($4$~s) for two initial 
conditions corresponding to $\omega=5$~Hz and $10$~Hz. 
The arrows indicate the increase of $\omega$, whereas the vertical 
bars indicate the uncertainty in correlations at the plateau values. 
Other parameters are identical as in Fig. \ref{gust_nu}. 
}
\label{corr_nu}
\end{figure}

After the Tonks-Girardeau gas expands in the disordered potential and reaches 
a steady-state, the behavior of $\rho_B(0,x,t)$ and $\mu_B(0,x,t)$ significantly 
differs from that at $t=0$. This is shown in Fig. \ref{rhot}, where we display the magnitude of the 
two functions for $\sigma=0.13$ and $\omega=10$~Hz. 
We observe that $|\rho_B(0,x,t)|$ exhibits a fairly fast exponential decay 
for small values of $|x|$, that is, in the region where the density is relatively 
large [see the inset in Fig. \ref{rhot}(a)].
This fast decay slows down up to sufficiently large values of $x$, i.e., $|x|>L_t$, 
where we observe slower exponential decay of $|\rho_B(0,x,t)|$, 
which corresponds to the exponentially decaying tails in the single-particle 
density of the localized steady state. 
Regarding the degree of first-order coherence $|\mu_B(0,x,t)|$, we find that 
for sufficiently small $|x|$, it decays exponentially 
[see the inset in Fig. \ref{rhot}(b)]; however, as $x$ approaches 
the region of exponentially decaying tails $|x|>L_t$, the exponential decay of $|\mu_B(0,x,t)|$ 
slows down until it reaches roughly a constant value in the region $|x|>L_t$. 
This plateau occurs because single-particle states $\psi_j$ decay slower for 
larger $j$ values (they are higher in energy and momentum), and due to the 
fact that for sufficiently large $|x|$, the matrix elements $A_{ij}(0,x,t)$, 
which are important ingredients in expression (\ref{expansion}) for $|\rho_B(0,x,t)|$, 
also reach a constant value. 
This is depicted in Figs. \ref{figpsij} and \ref{Aij}, which display $|\psi_j(x,t)|^2$ 
for $j=5,9$, and $13$, and $A_{1,13}(0,x,t)$ (real and imaginary part) for five different 
realizations of the disordered potential. We clearly see that $A_{ij}(0,x,t)$ 
reaches a constant value (generally complex off the diagonal), which differs from 
one realization of the disorder to the next; this is connected to the fact that the integral 
$\int_{0}^{x}dx' \psi_{i}^{*}(x',t)\psi_{j}(x',t)$ converges to a constant value 
for sufficiently large $x$, which is a consequence of the exponential localization. 
The fluctuations in $A_{ij}(0,x,t)$ are reflected onto the fluctuations of the 
plateau value of $|\mu_B(0,x,t)|$. 
We have compared the averages of the matrix elements $A_{ij}(0,x,t)$ for large 
$x$ (at the plateau) for all values of $i$ and $j$. They are all within one 
order of magnitude with $A_{13,13}(0,x,t)$ ($N=13$) being the largest, 
more specifically, the averages of some of the absolute value in our simulations are 
$|A_{13,13}(0,x,t)|=0.25\times 10^{-3}$, 
$|A_{7,7}(0,x,t)|=0.16\times 10^{-3}$, 
$|A_{1,1}(0,x,t)|=0.06\times 10^{-3}$, and 
$|A_{1,13}(0,x,t)|=0.03\times 10^{-3}$. 
Thus, the values of the matrix elements to some extent enhance the contribution of 
the highest single-particle states in the correlations $|\mu(0,x,t)|$. 
It is worthy to mention that identical effect is observed in incoherent light 
solitons (e.g., see \cite{IncOpt}), where the coherence also increases in the tails, 
which is observed in the complex coherence factor in optics (in the case of solitons, 
it is nonlinearity, rather than disorder which keeps the wave packet localized). 

Figure \ref{corr_sig} displays the correlations for three values of $\sigma$ ($\sigma=0.13$, 
$0.25$, and $0.40$). In the parameter regime that we investigated, we found no 
clear dependence of $|\rho_B(0,x,t)|$ and $|\mu_B(0,x,t)|$ (in the steady state 
at $4$~s) on $\sigma$ for small values of $|x|$. For $|x|$ in the region of the 
tails, $|x|>L_t$, the correlations $|\mu_B(0,x,t)|$ asymptote larger values for 
larger $\sigma$. 
All of the qualitative observations above were made throughout the parameter 
regime that we investigated numerically. 

In Figure \ref{corr_nu} we display $|\rho_B(0,x,t)|$ and $|\mu_B(0,x,t)|$, for two 
different initial conditions corresponding to $\omega=5$~Hz and $10$~Hz; other parameters are 
identical as in Fig. \ref{gust_nu}, that is, $\sigma=0.13$. 
We have observed that for asymptotic values of $t$, the magnitude of 
correlations $|\mu_B(0,x,t)|$ is lower for larger values of $\omega$. 
It is worthy to mention that a plateau in $|\mu_B(0,x,t)|$ occurs in 
every numerical simulation for a given realization of the disordered 
potential, but in every realization on a somewhat different value; the 
vertical bars in Figs. \ref{corr_sig} and \ref{corr_nu} indicate a spread 
in plateau values in our simulations. 

If we compare Fig. \ref{corr_sig} with Fig. \ref{gust_sigma}, and 
Fig. \ref{corr_nu} with Fig. \ref{gust_nu} (compare the direction of arrows 
in corresponding figures), one finds that if the density in the tails decays 
faster, the correlations between the center of the cloud and the tails decrease slower. 
This is in agreement with our interpretation that a slower decay (with $|x|$) of 
higher single-particle states $\psi_j$ (see Fig. \ref{figpsij}) is responsible for the 
creation of plateaus in $|\mu(0,x,t)|$ and increase of coherence in the 
tails; namely, if the density decays faster (with $|x|$), the highest single 
particle state $\psi_N$ will become dominant for smaller values of $x$ 
leading to greater coherence in $|\mu(0,x,t)|$. 

Let us now extrapolate our numerical calculations and results to larger 
particle numbers. Suppose that we keep all parameters fixed, and increase 
only $N$. The energy of the initial state as well as the high momentum cut-off 
$k_{hcm}$ increase with the increase of $N$. Our simulations up to a finite 
time up of $4$~s would not be able to see exponentially decaying tails
of the asymptotic steady state. By employing the results of Ref. 
\cite{Lugan2009}, one concludes that the steady state will always be 
localized, however, at larger values of $N$, the Born-approximation mobility 
edge \cite{Lugan2009} will be crossed and the exponents describing the 
exponentially decaying tails will be smaller. 
The plateaus in the correlations will still exist in the regions of these tails, 
however, the value $|\mu_B(0,x,t)|$ will decrease with the increase of $N$ (simply 
because more single particle states $\psi_j$ are needed to describe the 
Tonks-Girardeau state), and both the exponentially decaying tails together with 
the plateaus will be harder to observe. The effect where the coherence 
of the localized steady state increases in the tails should however be observable 
also with partially condensed BECs, below the Tonks-Girardeau regime. 

\section{Conclusion}
\label{Sec:Conc}

We have investigated Anderson localization of a Tonks-Girardeau gas in continuous 
potentials [$V_D(x)$] with controlled disorder, by investigating expansion of the 
gas in such potentials; for the initial state we have chosen the Tonks-Girardeau ground 
state in a harmonic trap (with $V_D(x)$ superimposed upon it), and we have analyzed 
the properties of the (asymptotic) steady state obtained dynamically. 
We have studied the dependence of the Lyapunov exponents and correlations on the 
initial trap parameter $\omega$ [$5-10$~Hz], and the correlation length of the 
disorder $\sigma$ [$0.13-0.40$~$\mu$m]. We found that the Lyapunov exponents of the 
steady state, decrease with the increase of $\nu$. 
In the parameter regime considered the Lyapunov exponents increased with the 
increase of $\sigma$, which was underpinned by the perturbation theory. 
The behavior of the correlations contained in the one-body density matrix 
$\rho_B(x,y,t)$ and the degree of first order coherence indicate that 
the off diagonal correlations $|\rho_B(0,x,t)|$ decrease exponentially with the increase of 
$|x|$, due to the exponential decay of the density, however, in the region of 
the exponentially decaying tails, the degree of first-order coherence 
$|\mu_B(0,x,t)|$ reaches a plateau. This is connected to the behavior of the 
single-particle states used to construct the Tonks-Girardeau wave function 
and to the increase of coherence in the exponentially decaying tails.
This effect is analogous to the one found in incoherent optical solitons, 
for which coherence also increases in the tails. 

As a possible direction for further research we envision a study of Anderson 
localization for incoherent light in disordered potentials, Anderson 
localization within the framework of the Lieb-Liniger model describing 
a 1D Bose gas with finite strength interactions (which 
becomes identical to the Tonks-Girardeau model when the interaction strength 
becomes infinite). These studies should provide further insight into the 
influence of wave coherence (within the context of optics), and the influence of 
interactions on Anderson localization (within the context of effectively 1D 
ultracold atomic gases). 

\acknowledgments
We are grateful to P. Lugan for helpful comments regarding the 
formalism of Ref. \cite{Lugan2009}. We are also grateful to the anonymous referee 
for suggesting to explore in more detail the role played by the matrix 
${\mathbf A}(x,y,t)$ in the correlations at the plateau. 
This work is supported by the Croatian-Israeli scientific collaboration, 
the Croatian Ministry of Science (Grant No. 119-0000000-1015), and 
the Croatian National Foundation of Science.

\begin{appendix}
\section{Construction of the disordered potential}
\label{app:VD}

In this section we describe the numerical procedure utilized for construction 
of the disordered potential $V_D(x)$. The $x$-space is numerically simulated 
by using 33000 equidistant points in the interval $x\in [-2000,2000]$. 
From this array, we have constructed a random 
array $v=\exp[2\pi\,\mbox{rand}(x)i]$ of the same length, where $\mbox{rand}(x)$ denotes a random 
number in between $0$ and $1$. 
Then we calculated a discrete Fourier transform of $v$ (call it $\tilde v(k)$), 
and introduced a cut-off wave vector $K_{cut}$. 
$V_D(x)$ was chosen to be an absolute value of the inverse discrete Fourier 
transform of $\tilde v(k) \Theta(2k/K_{cut})$ [where $\Theta(x)$ is one for $|x|<1$ 
and zero otherwise]. We have calculated the autocorrelation function 
$A_C(x)$ of the potential and fitted 
it to the functional form $\sin(x/\sigma)/(x/\sigma)^2$ to get the correlation 
length $\sigma$.
The autocorrelation function $A_C(x)$ of the disordered potential $V_D(x)$
is identical to the autocorrelation function of the potential used in the experiment 
of Ref. \cite{Billy2008}, and the theoretical studies conducted in 
Refs. \cite{Sanchez-Palencia2007,Lugan2009}. 
The higher order correlators differ, but they do not qualitatively change any 
conclusions in the parameter regime studied here. 

Let us define the term $\gamma^{(3)}(k)$ which is used in the series 
$\gamma(k)=\sum_{n\geq 2}\gamma^{(n)}$ (see Ref. \cite{Lugan2009} and 
Fig. \ref{gamma_sigma}). This term depends on the 3-point correlator of the 
random potential \cite{Lugan2009}:
\begin{equation}
c_3(\frac{x_1}{\sigma},\frac{x_2}{\sigma})=\frac {1}{V_0^3} \ \langle \overline{V}_{D}(x'-x_1) 
\overline{V}_{D}(x'-x_2) \overline{V}_{D}(x') \rangle_{x'},
\end{equation}
and it is given by:
\begin{equation}
\gamma^{(3)}(k)=\frac{2{V_{0}}^{3} \sigma^2}{k^3} f_3(k \sigma),
\end{equation}
where $f_3(\kappa)$ is defined as:
\begin{equation}
f_3(\kappa)=-\frac{1}{4}\int^0_{-\infty}\, \! \! du\int^u_{-\infty} dv \ c_3(u,v)\sin(2 \kappa v)\ .
\end{equation}
The 3-point correlator was calculated numerically to obtain Fig. \ref{gamma_sigma}.

\end{appendix}



\begin{thebibliography}{99}

\bibitem{Anderson1958}
P.W. Anderson, 
Phys. Rev. {\bf 109}, 1492 (1958).


\bibitem{Wiersma1997}
D.S. Wiersma, P. Bartolini, A. Lagendijk, R. Righini, 
Nature {\bf 390}, 671 (1997).

\bibitem{Chabanov2000}
A.A. Chabanov, M. Stoytchev, A.Z. Genack, 
Nature {\bf 404}, 850 (2000).

\bibitem{Storzer2006}
M. St\" orzer, P. Gross, C.M. Aegerter, G. Maret, 
Phys. Rev. Lett. {\bf 96}, 063904 (2006).

\bibitem{Schwartz2007}
T. Schwartz, G. Bartal, S. Fishman, and M. Segev,
Nature {\bf 446}, 52 (2007). 

\bibitem{Lahini2008}
Y. Lahini, A. Avidan, F. Pozzi, M. Sorel, R. Morandotti,
D.N. Christodoulides, and Y. Silberberg, 
Phys. Rev. Lett. {\bf 100}, 013906 (2008). 


\bibitem{Billy2008}
J. Billy, V. Josse, Z. Zuo, A. Bernard, B. Hambrecht, P. Lugan, 
D. Clement, L. Sanchez-Palencia, P. Bouyer, A. Aspect,
Nature {\bf 453}, 891 (2008). 

\bibitem{Roati2008}
G. Roati, C. D'Errico, L. Fallani, M. Fattori, C. Fort, M. Zaccanti,
G. Modugno, M. Modugno, M. Inguscio,
Nature {\bf 453}, 895 (2008). 



\bibitem{Sanchez-Palencia2010}
L. Sanchez-Palencia and M. Lewenstein, 
Nature Phys. {\bf 6}, 87 (2010). 


\bibitem{Girardeau1960}
M. Girardeau,
J. Math. Phys. {\bf 1}, 516 (1960). 

\bibitem{Girardeau2000}
M. Girardeau and E.M. Wright,
Phys. Rev. Lett. {\bf 84}, 5691 (2000). 


\bibitem{Olshanii98}
M. Olshanii, Phys. Rev. Lett. {\bf 81}, 938 (1998).

\bibitem{Petrov2000}
D.S. Petrov, G.V. Shlyapnikov, and J.T.M. Walraven,
Phys. Rev. Lett. {\bf 85} 3745 (2000).

\bibitem{Dunjko2001}
V. Dunjko, V. Lorent, and M. Olshanii,
Phys. Rev. Lett. {\bf 86} 5413 (2001).


\bibitem{Kinoshita2004}
T. Kinoshita, T. Wenger, and D.S. Weiss,
Science {\bf 305}, 1125 (2004). 

\bibitem{Paredes2004}
B. Paredes, A. Widera, V. Murg, O. Mandel, S. F\" olling, 
I. Cirac, G. V. Shlyapnikov, T. W. H\" ansch, and I. Bloch, 
Nature (London) {\bf 429}, 277 (2004). 

\bibitem{Kinoshita2006}
T. Kinoshita, T. Wenger, and D.S. Weiss,
Nature (London) {\bf 440}, 900 (2006). 


\bibitem{Lenard1964}
A. Lenard, J. Math. Phys. {\bf 5}, 930 (1964).

\bibitem{Forrester2003}
P.J. Forrester, N.E. Frankel, T.M. Garoni, and N.S. Witte,
Phys. Rev. A {\bf 67}, 043607 (2003); T. Papenbrock,
Phys. Rev. A {\bf 67}, 041601 (2003).


\bibitem{Rigol2005exp}
M. Rigol and A. Muramatsu,
Phys. Rev. Lett {\bf 94}, 240403 (2005). 

\bibitem{Minguzzi2005}
A. Minguzzi and D.M. Gangardt,
Phys. Rev. Lett. {\bf 94}, 240404 (2005).

\bibitem{DelCampo2006}
A. del Campo and J.G. Muga,
Europhys. Lett. {\bf 74}, 965 (2006).

\bibitem{Gangardt2007}
D.M. Gangardt and M. Pustilnik,
Phys. Rev. A {\bf 77}, 041604(R) (2008).


\bibitem{Damski2003}
B. Damski, J. Zakrzewski, L. Santos, P. Zoller, and M. Lewenstein,
Phys. Rev. Lett. {\bf 91}, 080403 (2003).

\bibitem{Roth2003}
R. Roth and K. Burnett,
Phys. Rev. A {\bf 68}, 023604 (2003). 

\bibitem{Sanchez-Palencia2007}
L. Sanchez-Palencia, D. Clement, P. Lugan, P. Bouyer, 
G.V. Shlyapnikov, and A. Aspect,
Phys. Rev. Lett. {\bf 98}, 210401 (2007).

\bibitem{Giamarchi1988}
T. Giamarchi and H.J. Schulz, 
Phys. Rev. B {\bf 37}, 325 (1988).

\bibitem{Fisher1989}
M.P.A. Fisher, P.B. Weichman, G. Grinstein, and D.S. Fisher, 
Phys. Rev. B {\bf 40}, 546 (1989).

\bibitem{Gimperlein2005}
H. Gimperlein, S. Wessel, J. Schmiedmayer, and L. Santos,
Phys. Rev. Lett. {\bf 95}, 170401 (2005).

\bibitem{deMartino2005}
A. De Martino, M. Thorwart, R. Egger, and R. Graham,
Phys. Rev. Lett. {\bf 94}, 060402 (2005). 

\bibitem{Scarola2006}
V.W. Scarola and S. Das Sarma, 
Phys. Rev. A {\bf 73}, 041609(R) (2006).

\bibitem{Rey2006}
A.M. Rey, I.I. Satija, and C.W. Clark, 
Phys. Rev. A {\bf 73}, 063610 (2006).

\bibitem{Horstmann2007}
B. Horstmann, J.I. Cirac, and T. Roscilde,
Phys. Rev. A {\bf 76}, 043625 (2007). 

\bibitem{Roux2008}
G. Roux, T. Barthel, I.P. McCulloch, C. Kollath, U. Schollw\" ock, and T. Giamarchi,
Phys. Rev. A {\bf 78}, 023628 (2008). 

\bibitem{Deng2008}
X. Deng, R. Citro, A. Minguzzi, and E. Orignac, 
Phys. Rev. A, {\bf 78}, 013625 (2008).

\bibitem{Roscilde2008}
T. Roscilde, 
Phys. Rev. A {\bf 77}, 063605 (2008).

\bibitem{Orso2009}
G. Orso, A. Iucci, M.A. Cazalilla, and T. Giamarchi,
Phys. Rev. A {\bf 80}, 033625 (2009). 


\bibitem{Rigol2005}
M. Rigol and A. Muramatsu,
Phys. Rev. A {\bf 72}, 013604 (2005); 
{\it ibid.}, {\bf 70}, 031603 (2004).


\bibitem{IncOpt}
M. Mitchell, M. Segev, T.H. Coskun, and D.N.
Christodoulides, Phys. Rev. Lett. {\bf 79}, 4990 (1997);
M.I. Carvalho, T.H. Coskun, D.N. Christodoulides, M. Mitchell, and M. Segev,
Phys. Rev. E {\bf 59}, 1193 (1999); 
H. Buljan, T. Schwartz, M. Segev, M. Solja\v{c}i\'{c}, and D.N. Christodoulides, 
J. Opt. Soc. Am. B {\bf 21}, 397 (2004). 


\bibitem{Buljan2006}
H. Buljan, O. Manela, R. Pezer, A. Vardi, and M. Segev,
Phys. Rev. A {\bf 74}, 043610 (2006).

\bibitem{Pezer2007}
R. Pezer and H. Buljan, Phys. Rev. Lett. {\bf 98}, 240403 (2007).


\bibitem{Lugan2009}
P. Lugan, A. Aspect, and L. Sanchez-Palencia, D. Delande, B. Gremaud, 
C.A. M\" uller, C. Miniatura, Phys. Rev. A {\bf 80}, 023605 (2009).

\bibitem{notation}
We utilize notation from Ref. \cite{Lugan2009} to denote the orders of 
the perturbation, however, the coefficients $\gamma^{(n)}$ here describe the 
decay of density, rather than the wave function as in \cite{Lugan2009}, and 
they differ by a factor of $2$. 

\bibitem{Naraschewski1999}
M. Naraschewski and R.J. Glauber, 
Phys. Rev. A {\bf 59}, 4595 (1999). 

\bibitem{MandelWolf}
L. Mandel and E. Wolf, 
{\it Optical Coherence and Quantum Optics}
(Cambridge Press, New York, 1995).

\end{thebibliography}
\end{document}